\documentclass[%
reprint,
superscriptaddress,
 amsmath,amssymb,
 aps,
 prl
]{revtex4-1}

\usepackage{color}
\usepackage{graphicx}
\usepackage{dcolumn}
\usepackage{bm}

\begin{document}

\preprint{CLAS Collaboration}

\title{ Photoproduction of the $f_2(1270)$ meson using the CLAS detector}

\newcommand*{\OHIOU}{Ohio University, Athens, Ohio  45701}
\newcommand*{\OHIOUindex}{1}
\newcommand*{\INFNGE}{INFN, Sezione di Genova, 16146 Genova, Italy}
\newcommand*{\INFNGEindex}{2}
\newcommand*{\UCM}{Departamento de Física Teórica and IPARCOS, Universidad Complutense de Madrid, 28040 Madrid, Spain}
\newcommand*{\UCMindex}{3}
\newcommand*{\NOWISU}{Idaho State University, Pocatello, Idaho 83209}
\newcommand*{\ECT}{European Centre for Theoretical Studies in Nuclear Physics and Related Areas (ECT$^*$) and Fondazione Bruno Kessler, Strada delle Tavarnelle 286, Villazzano (Trento), I-38123 Italy}

\newcommand*{\ANL}{Argonne National Laboratory, Argonne, Illinois 60439}
\newcommand*{\ANLindex}{4}
\newcommand*{\ASU}{Arizona State University, Tempe, Arizona 85287-1504}
\newcommand*{\ASUindex}{5}
\newcommand*{\CMU}{Carnegie Mellon University, Pittsburgh, Pennsylvania 15213}
\newcommand*{\CMUindex}{6}
\newcommand*{\CUA}{Catholic University of America, Washington, D.C. 20064}
\newcommand*{\CUAindex}{7}
\newcommand*{\SACLAY}{IRFU, CEA, Universit\'{e} Paris-Saclay, F-91191 Gif-sur-Yvette, France}
\newcommand*{\SACLAYindex}{8}
\newcommand*{\CNU}{Christopher Newport University, Newport News, Virginia 23606}
\newcommand*{\CNUindex}{9}
\newcommand*{\UCONN}{University of Connecticut, Storrs, Connecticut 06269}
\newcommand*{\UCONNindex}{10}
\newcommand*{\DUKE}{Duke University, Durham, North Carolina 27708-0305}
\newcommand*{\DUKEindex}{11}
\newcommand*{\DUQUESNE}{Duquesne University, 600 Forbes Avenue, Pittsburgh, PA 15282 }
\newcommand*{\DUQUESNEindex}{12}
\newcommand*{\FU}{Fairfield University, Fairfield CT 06824}
\newcommand*{\FUindex}{13}
\newcommand*{\FERRARAU}{Universita' di Ferrara , 44121 Ferrara, Italy}
\newcommand*{\FERRARAUindex}{14}
\newcommand*{\FIU}{Florida International University, Miami, Florida 33199}
\newcommand*{\FIUindex}{15}
\newcommand*{\FSU}{Florida State University, Tallahassee, Florida 32306}
\newcommand*{\FSUindex}{16}
\newcommand*{\GWUI}{The George Washington University, Washington, DC 20052}
\newcommand*{\GWUIindex}{17}
\newcommand*{\ISU}{Idaho State University, Pocatello, Idaho 83209}
\newcommand*{\ISUindex}{18}
\newcommand*{\INFNFE}{INFN, Sezione di Ferrara, 44100 Ferrara, Italy}
\newcommand*{\INFNFEindex}{19}
\newcommand*{\INFNFR}{INFN, Laboratori Nazionali di Frascati, 00044 Frascati, Italy}
\newcommand*{\INFNFRindex}{20}
\newcommand*{\INFNRO}{INFN, Sezione di Roma Tor Vergata, 00133 Rome, Italy}
\newcommand*{\INFNROindex}{21}
\newcommand*{\INFNTUR}{INFN, Sezione di Torino, 10125 Torino, Italy}
\newcommand*{\INFNTURindex}{22}
\newcommand*{\INFNPAV}{INFN, Sezione di Pavia, 27100 Pavia, Italy}
\newcommand*{\INFNPAVindex}{23}
\newcommand*{\ORSAY}{Universit'{e} Paris-Saclay, CNRS/IN2P3, IJCLab, 91405 Orsay, France}
\newcommand*{\ORSAYindex}{24}
\newcommand*{\Juelich}{Institute fur Kernphysik (Juelich), Juelich, Germany}
\newcommand*{\Juelichindex}{25}
\newcommand*{\JMU}{James Madison University, Harrisonburg, Virginia 22807}
\newcommand*{\JMUindex}{26}
\newcommand*{\KNU}{Kyungpook National University, Daegu 41566, Republic of Korea}
\newcommand*{\KNUindex}{27}
\newcommand*{\LAMAR}{Lamar University, 4400 MLK Blvd, PO Box 10046, Beaumont, Texas 77710}
\newcommand*{\LAMARindex}{28}
\newcommand*{\MISS}{Mississippi State University, Mississippi State, MS 39762-5167}
\newcommand*{\MISSindex}{29}
\newcommand*{\ITEP}{National Research Centre Kurchatov Institute - ITEP, Moscow, 117259, Russia}
\newcommand*{\ITEPindex}{30}
\newcommand*{\UNH}{University of New Hampshire, Durham, New Hampshire 03824-3568}
\newcommand*{\UNHindex}{31}
\newcommand*{\NSU}{Norfolk State University, Norfolk, Virginia 23504}
\newcommand*{\NSUindex}{32}
\newcommand*{\ODU}{Old Dominion University, Norfolk, Virginia 23529}
\newcommand*{\ODUindex}{33}
\newcommand*{\JLUGiessen}{II Physikalisches Institut der Universitaet Giessen, 35392 Germany}
\newcommand*{\JLUGiessenindex}{34}
\newcommand*{\ROMAII}{Universita' di Roma Tor Vergata, 00133 Rome Italy}
\newcommand*{\ROMAIIindex}{35}
\newcommand*{\MSU}{Skobeltsyn Institute of Nuclear Physics, Lomonosov Moscow State University, 119234 Moscow, Russia}
\newcommand*{\MSUindex}{36}
\newcommand*{\SCAROLINA}{University of South Carolina, Columbia, South Carolina 29208}
\newcommand*{\SCAROLINAindex}{37}
\newcommand*{\TEMPLE}{Temple University,  Philadelphia, PA 19122 }
\newcommand*{\TEMPLEindex}{38}
\newcommand*{\JLAB}{Thomas Jefferson National Accelerator Facility, Newport News, Virginia 23606}
\newcommand*{\JLABindex}{39}
\newcommand*{\RICH}{University of Richmond, Richmond, Virginia 23173}
\newcommand*{\RICHindex}{40}
\newcommand*{\UTFSM}{Universidad T\'{e}cnica Federico Santa Mar\'{i}a, Casilla 110-V Valpara\'{i}so, Chile}
\newcommand*{\UTFSMindex}{41}
\newcommand*{\INSUBRIA}{Universit\`{a} degli Studi dell'Insubria, 22100 Como, Italy}
\newcommand*{\INSUBRIAindex}{42}
\newcommand*{\BRESCIA}{Universit\`{a} degli Studi di Brescia, 25123 Brescia, Italy}
\newcommand*{\BRESCIAindex}{43}
\newcommand*{\GLASGOW}{University of Glasgow, Glasgow G12 8QQ, United Kingdom}
\newcommand*{\GLASGOWindex}{44}
\newcommand*{\YORK}{University of York, York YO10 5DD, United Kingdom}
\newcommand*{\YORKindex}{45}
\newcommand*{\VT}{Virginia Tech, Blacksburg, Virginia   24061-0435}
\newcommand*{\VTindex}{46}
\newcommand*{\VIRGINIA}{University of Virginia, Charlottesville, Virginia 22901}
\newcommand*{\VIRGINIAindex}{47}
\newcommand*{\WM}{College of William and Mary, Williamsburg, Virginia 23187-8795}
\newcommand*{\WMindex}{48}
\newcommand*{\YEREVAN}{Yerevan Physics Institute, 375036 Yerevan, Armenia}
\newcommand*{\YEREVANindex}{49}


\author{M.~Carver}
\affiliation{\OHIOU}
\author{A.~Celentano}
\affiliation{\INFNGE}
\author{K.~Hicks}
\affiliation{\OHIOU}
\author {L.~Marsicano} 
\affiliation{\INFNGE}
\author{V.~Mathieu}
\affiliation{\UCM}
\author{A.~Pilloni}
\affiliation{\ECT}
\affiliation{\INFNGE}
\affiliation{\INFNRO}
\author {K.P.~Adhikari} 
\affiliation{\ODU}
\author {S. Adhikari} 
\affiliation{\FIU}
\author {M.J.~Amaryan} 
\affiliation{\ODU}
\author {Giovanni Angelini} 
\affiliation{\GWUI}
% ...

\collaboration{The CLAS Collaboration}
\noaffiliation

\date{\today}

\begin{abstract}
The quark structure of the $f_2(1270)$ meson has, for many years, been assumed to be a pure quark-antiquark ($q\bar{q}$) resonance with quantum numbers $J^{PC} = 2^{++}$. Recently, it was proposed that the $f_2(1270)$ is a molecular state made from the attractive interaction of two $\rho$-mesons.  Such a state would be expected to decay strongly to final states with charged pions, due to the dominant decay $\rho \to \pi^+ \pi^-$, whereas decay to two neutral pions would likely be suppressed.  Here, we measure for the first time the reaction $\gamma p \to \pi^0 \pi^0 p$, using the CLAS detector at Jefferson Lab for incident beam energies between 3.6-5.4~GeV.  Differential cross sections, $d\sigma / dt$, for $f_2(1270)$ photoproduction are extracted with good precision, due to low backgrounds, and are compared with theoretical calculations. 
\end{abstract}

\maketitle


There are several possible models in the literature for the internal structure of the tensor meson $f_2(1270)$. In the standard quark model \cite{Tanabashi}, it is a simple $q\bar{q}$ pair with spins aligned, $S=1$, and one unit of orbital angular momentum, $L=1$. In spectroscopic notation, it is a $^3P_2$ state, with $J=2$. The quark model groups particles of similar total spin $J$ and parity $P$ together, so the $f_2(1270)$ is the isosinglet in a nonet group that includes the $a_2(1320), K^*(1430)$ and $f_2'(1525)$ mesons.

A different model, where the $f_2(1270)$ is a resonance dynamically generated from the interaction of two $\rho$-mesons, was introduced by Molina {\it et al.} \cite{Molina}. Using this model, Ref. \cite{Xie} calculated the photoproduction cross section of the $f_2(1270)$ decaying to $\pi^+ \pi^-$ and compared it to the CLAS data \cite{Marco} even though that comparison was indirect (as explained below). This model has few free parameters, which are mostly constrained by other data, and so the agreement between theory and experiment offered an alternative explanation of the $f_2(1270)$ structure as a $\rho$-$\rho$ molecule.

A third possibility is that the $f_2(1270)$ mixes with the lowest-mass tensor glueball \cite{Yu}, both having the same $J^{PC}=2^{++}$.  This model is based on ratios of the decay of $J/\psi$ and $\psi'$ to the $\gamma + f_2(1270)$ final state. This suggestion of glueball mixing in the $f_2(1270)$ structure has been contested by some authors \cite{DeMinLi}, but a small mixing is still plausible in an effective field approach \cite{Giacosa}. 

These differing ideas for the $f_2(1270)$ structure motivate the need for more data starting from a simple initial state such as the photoproduction reaction $\gamma p \to f_2(1270) p$. Here we report on this reaction from the $g12$ experiment, using the CLAS detector \cite{mecking}.

The reaction $\gamma p \rightarrow f_2(1270) p \rightarrow \pi^0 \pi^0 p$ is an excellent channel to investigate the $f_2(1270)$ resonance, since unlike the $\pi^+ \pi^-$ decay channel, there is no $\rho$ meson signal. Therefore, extracting the $f_2(1270)$ signal becomes easier, as it avoids large backgrounds.  Given the indistinguishability of the two neutral mesons in the final state, Bose-Einstein statistical rules act as a $J^{PC}$ filter, allowing only even-$L$ partial waves to contribute to the final state. This removes the dominant $\rho$ background that characterized past studies using the $\pi^+ \pi^-$ final state. There are no published cross sections for $f_2(1270)$ production from the $\gamma p \to \pi^0 \pi^0 p$ reaction at small momentum transfers, where theoretical models based on Regge exchange are applicable. 

The first published analysis on the $f_2(1270)$ meson was in 1976 \cite{Clifft}. That paper investigated the $\pi^+ \pi^-$ channel, which has a significant contribution from the $\rho$ meson. For the event yield extraction, all counts between 1100 and 1400 MeV were taken as belonging to the $f_2(1270)$ meson. Therefore, their event yield for the $f_2(1270)$ includes some of the $\rho$ meson background. In 2009, the CLAS Collaboration measured the $f_2(1270)$ \cite{Marco} via its $\pi^+ \pi^-$ decay, integrated over photon beam energies from 3.0 to 3.8~GeV. There, the $D$-wave part of the cross section was extracted in the presence of a large $\rho$-meson background by using a partial wave analysis (PWA), which had large uncertainties (error bars of $\sim$40\%).  A recent theoretical paper \cite{mathieu} based on Regge theory used these $D$-wave results to extract the $f_2(1270)$ cross sections, which were compared with two models.  These models are compared to the new results below.

The present analysis utilizes a tagged photon beam \cite{tagger} with energy range 3.6 to 5.4~GeV on a 40-cm-long liquid-hydrogen target, leading to the reaction $\gamma p \to \pi^0 \pi^0 p$. The goal of this analysis is to learn about the structure of the $f_2(1270)$ through comparison of theoretical models with the experimental cross section $d\sigma / dt$, where $t$ is the four-momentum transfer squared between the beam photon and the outgoing proton. Since $\rho$ decay to $2\pi^0$ is forbidden, a clean $f_2(1270)$ signal is seen in the $\pi^0\pi^0$ invariant mass spectrum, enabling fine binning of the cross section for the incident beam energy as a function of $t$.  

Data from the $g12$ experiment \cite{g12} were collected in the spring of 2008 with the CEBAF Large Acceptance Spectrometer (CLAS) \cite{mecking} at the Thomas Jefferson National Accelerator Facility. The CLAS detector had six superconducting coils that produced a toroidal field around the beam direction. Six sets of drift chambers (DC) determined the charged-particle trajectories, with gas Cherenkov counters to distinguish electrons and pions, plastic scintillator bars to measure the time-of-flight (TOF), and an electromagnetic calorimeter (EC) to detect neutrals and electrons. 
A plastic scintillator hodoscope (ST) surrounded the target to measure the start time. A high-speed data acquisition system read out the detector system.
The photon beam flux was $\sim 10^7$/s. 

The main trigger condition for the $g12$ experiment required the presence of one charged particle, defined as a coincidence between one TOF hit and one ST hit in the same CLAS sector, and two final-state photons in different CLAS sectors, each defined as an EC hit above a threshold of approximately 100~MeV.
The efficiency of the trigger system was evaluated from special minimum bias runs and found to be on average $\varepsilon_\text{trg}=83\%$. To account for the trigger efficiency dependence on the proton impact point on the detector, a trigger efficiency map, as a function of the proton three-momentum, was used for small corrections to the cross-section normalization.

The data were filtered to select events that had four neutral hits in the EC above a photon-energy threshold. One positively charged track was identified as a proton, using the DC for its trajectory and the TOF to get its speed. The tagged beam photon was selected to be within 1.0 ns of the proton's vertex time. 
Only events with exactly one tagged photon satisfying this criteria were further considered. These corresponded to a fraction $f_{1\gamma}=86.5\%$. The final event yield was corrected by a factor $1/f_{1\gamma}$ to account for this effect. 
Fiducial cuts on the active volume of the EC were applied to the four final state photons, and a vertex cut was applied to ensure the proton's track originated from the target volume.  A complete simulation of the CLAS detector was performed to obtain the detection efficiency (or acceptance) of the desired final state. The same analysis algorithm was used for both data and Monte Carlo.  Comparison of simulations (see below) and data corrected for a small ($\sim 9$\%) loss of the recoil proton detection probability in the ST. 

\begin{figure}
    \centering
    \includegraphics[width=.5\textwidth]{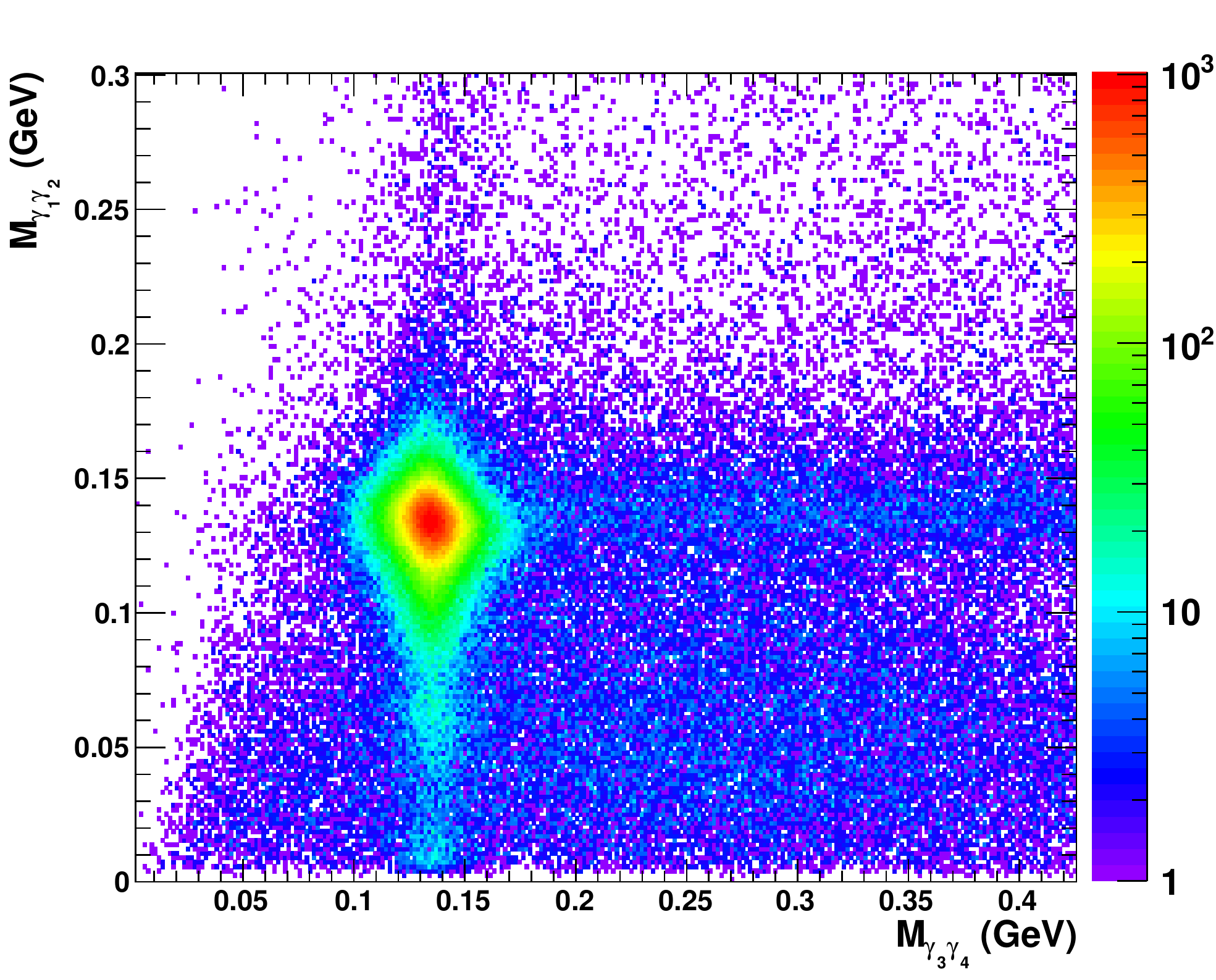}
    \caption{\label{fig:ordered}Correlation between the invariant mass of the two photon pairs for exclusive $\gamma p \rightarrow 4\gamma \,p$ events. 
    In each event, $\gamma_1$ and $\gamma_2$ are the photons with the smallest opening angle.
    The bottom-left cluster contains signal events from the $\gamma p \rightarrow \pi^0 \pi^0 p$ reaction.}
    \label{fig:my_label}
\end{figure}

The first part of the analysis was based on the same procedures for the recent CLAS analysis of the $\gamma p \rightarrow \pi^0 \eta p$ reaction described in Ref.~\cite{AndreaEtaPi}. 
A 4C kinematic fit (four constraints, imposing energy and momentum conservation)
was used to select events belonging to the exclusive $\gamma p \rightarrow 4\gamma p$ reaction, by introducing a cut on the corresponding confidence level (CL). The kinematic fit was tuned to the detector resolution to ensure a flat confidence-level (CL) distribution above about 20$\%$ CL.  Events with CL$<$10$\%$ were rejected in both data and Monte Carlo.  The result was a clean sample of exclusive events dominated by the $\pi^0 \pi^0 p$ final state.

The following procedure was then adopted to isolate the $\gamma p\rightarrow \pi^0 \pi^0 p$ reaction \cite{clasNotes}. First, the photons were ordered event-by-event by naming $\gamma_1$ and $\gamma_2$ those with the smallest opening angle; the other pair being named $\gamma_3$ and $\gamma_4$. This algorithm exploits the fact that, due to the low pion mass and to the Lorentz boost, two photons originating from the same $\pi^0$ are expected to have a smaller relative angle compared to two $\gamma$ from different parent particles. After ordering the photons, the $M_{\gamma_3 \gamma_4}$ and the $M_{\gamma_1 \gamma_2}$ distributions showed a clear peak corresponding to the $\pi^0\pi^0$ topology. The result is reported in Fig. \ref{fig:ordered}, showing the correlation between the invariant masses of the two photon pairs, $M_{\gamma_1 \gamma_2}$ vs. $M_{\gamma_3 \gamma_4}$.  
A very clear $\pi^0\pi^0$ signal is present, over a small background. The clean signal is a result of an EC threshold cut, along with the CL cut and the coincidence timing requirements.



\begin{figure}[htb]
\includegraphics[width=.45\textwidth]{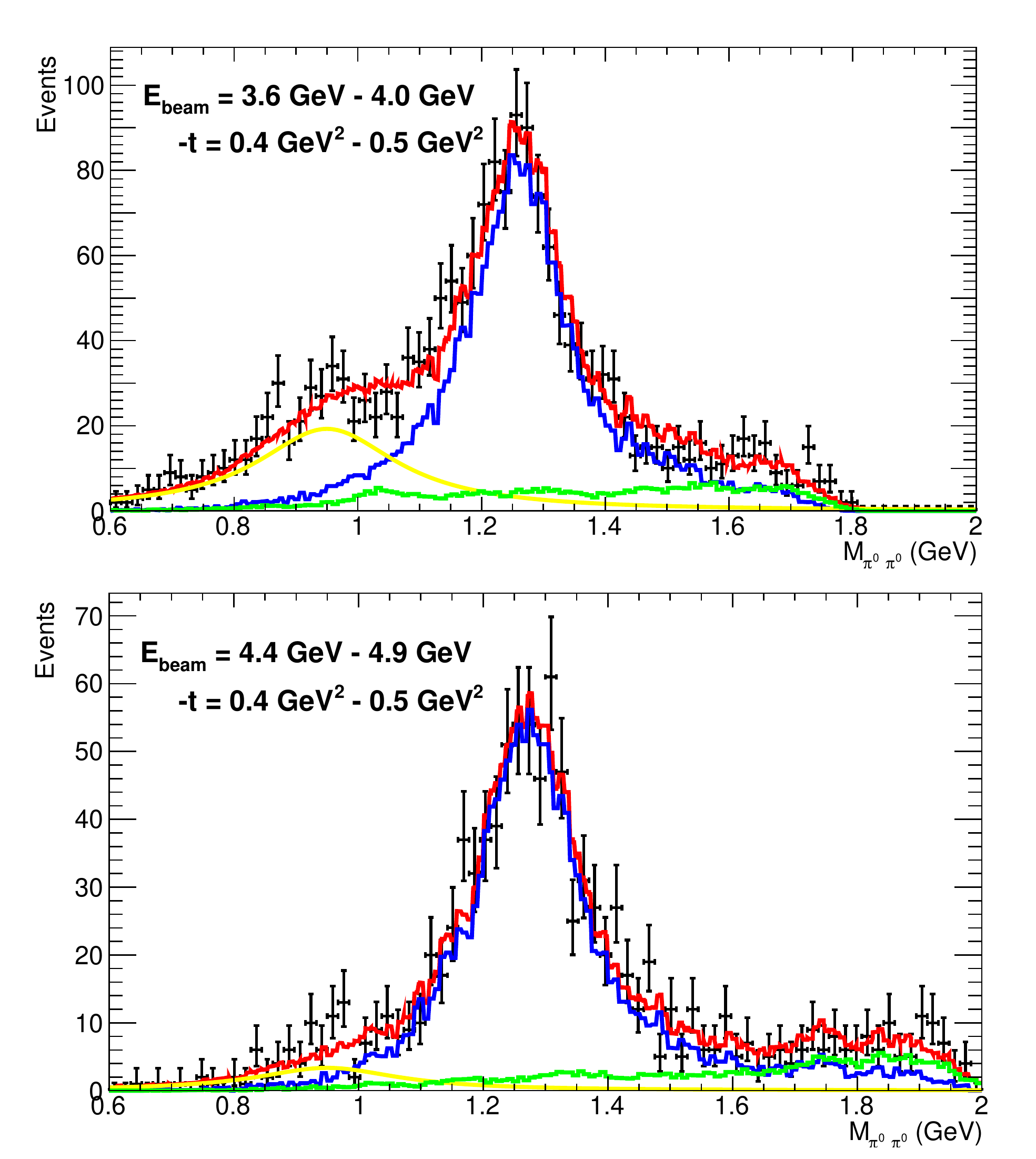}
\caption{\label{fig:f2mass}
Result of the maximum likelihood binned fit to the $\pi^0 \pi^0$ invariant mass distribution for two representative bins, as reported in the panels. The red curve is the full fit PDF, while the blue, green, and yellow curves represent, respectively, the $f_2$ signal PDF, the phase-space background PDF, and the low-mass background PDF.}
\end{figure}

The two-photons invariant mass distributions were fit with a Gaussian function to determine the width of the $\pi^0$ peak. After requiring that each $2\gamma$ invariant mass be within $\pm 3 \sigma$ of the $\pi^0$ mass, the data was divided into bins of the tagged photon energy $E_\gamma$ and the squared 4-momentum transfer to the proton, $t$. Then the $\pi^0\pi^0$ invariant mass was calculated for each event in a given bin.  

The $f_2(1270)$ event yield was extracted as follows \cite{clasNotes}. An extended maximum likelihood binned fit was performed to all invariant mass distributions, using a Probability Density Function (PDF) modeled as the incoherent sum of a signal term for the $f_2(1270)$ meson, and two background terms, one for the invariant mass range below the peak (in the region of the $f_0(980)$ meson) and the other for the range above the peak where incoherent (phase-space) production occurs. The $f_2(1270)$ event yield in each bin was then obtained as the integral of the signal term.
The signal PDF was obtained by simulating the $\gamma p \rightarrow f_2p$ reaction, with the resonance line-shape taken as a Breit-Wigner function with a mass of 1.26~GeV and a width of 0.183~GeV. The resonance mass and width were varied simultaneously in all bins to obtain the best fit, and are consistent with the values found by the Particle Data Group \cite{Tanabashi}. 
One bin, at the lowest $E_\gamma$ and $-t=0.15$ GeV$^2$, gave an unacceptable fit and was thus removed from our sample. A fit example is reported in Fig.~\ref{fig:f2mass}, showing the $\pi^0 \pi^0$ invariant mass distribution and the fit result for two different kinematic bins. The red curve is the full fit PDF, while the blue, green, and yellow curves represent, respectively, the $f_2$ signal PDF, the phase-space background PDF, and the low-mass background PDF.

A custom event generator was used to produce Monte Carlo events for this reaction, which were passed through a realistic detector simulation and the same reconstruction chain as for the data. The invariant mass distribution of reconstructed Monte Carlo events, for the same $E_\gamma$ and $t$ bins, was then used to derive the template for the signal PDF. A similar procedure was adopted for the high-mass background, which was obtained from a pure 3-particle phase-space distribution. Finally, the low-mass background was effectively parameterized with a Breit-Wigner function, centered at the $f_0(980)$ nominal mass \cite{Tanabashi}.
Additional fits were done by adding a template for the $f_0(1370)$, using the PDG values \cite{Tanabashi} for its mass and width, but this changed the fits only by a few percent in a few bins at high $E_\gamma$ and high $-t$, leaving most $f_2(1270)$ yields nearly the same (within 1\%). The systematic uncertainty associated with the fitting procedure was estimated at 4$\%$.

The CLAS detector acceptance was modeled using a computer program, {\it GSIM}, based on the GEANT software \cite{geant3}. After applying the same cuts as in the data analysis, the acceptance of the $\pi^0 \pi^0 p$ final state ranged between 0.4\% and 2.2\% for all kinematic bins. The acceptance was lowest for $E_\gamma > 5.0$~GeV and $-t < 0.3$~GeV$^2$.  From variations in the $t$-dependence of the $f_2(1270)$ event generator, we attribute a systematic uncertainty of 3\% to the detector acceptance. 

The largest source of systematic uncertainty was the beam flux, which was reported in detail in a previous paper from the $g12$ experiment \cite{kunkel}, with an uncertainty of 6\%.  Other sources of systematic uncertainties include the variation of kinematic cuts (3\%), target properties (1\%), $f_{1\gamma}$ correction (0.9$\%$), and branching ratios ($<$1\%). The overall systematic uncertainty is estimated at 8-10\%, depending slightly on the kinematic bin.

\begin{figure}[t]
\includegraphics[scale=0.45]{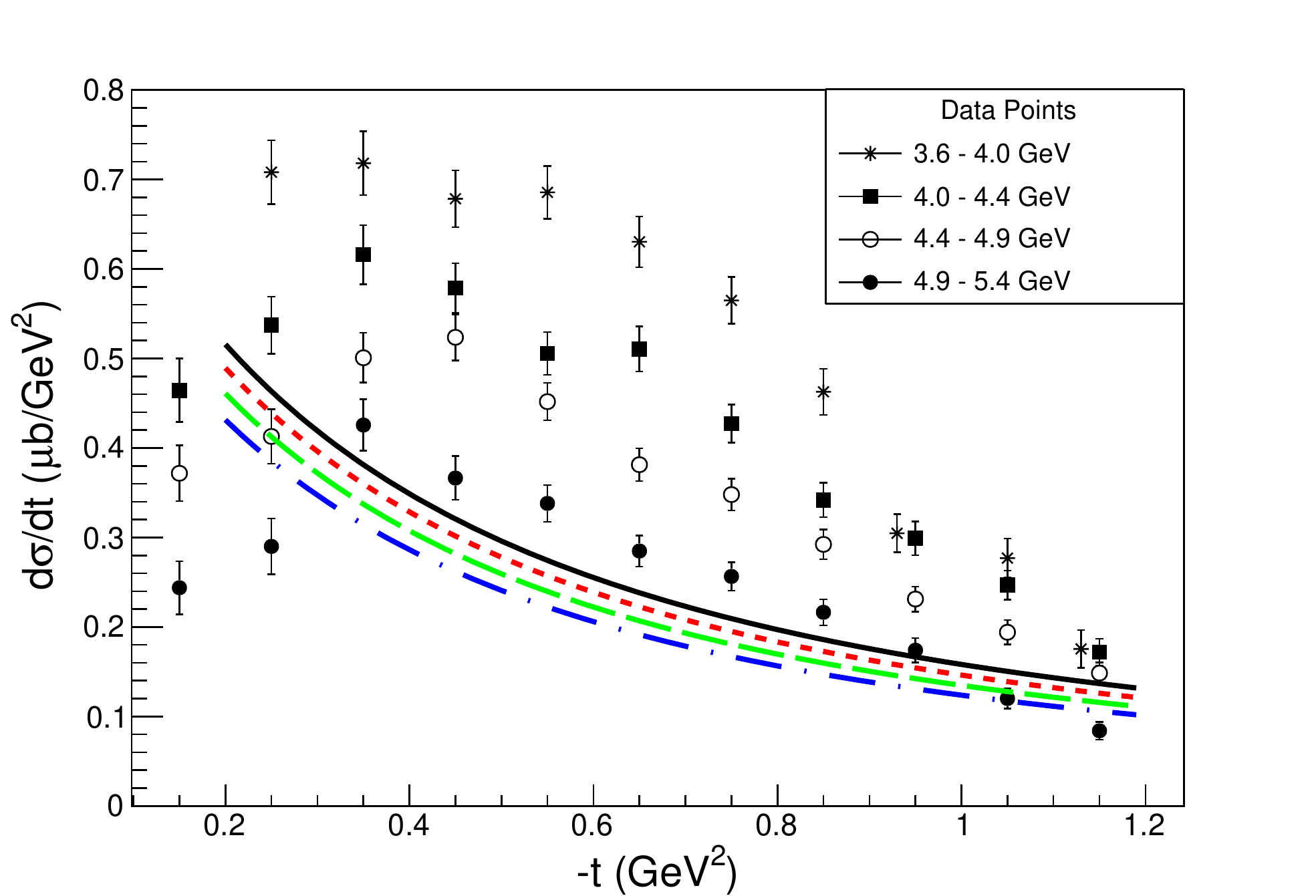}
\caption{\label{fig:xsec} Cross sections for the reaction $\gamma p \to f_2(1270) p$ as a function $-t$ for the given beam energies.  Two points at the lowest beam energy are slightly offset from the center of the $t$-bin for visibility. The curves are from model A of Xie and Oset \cite{Xie}.  See also the legend of Fig.~\ref{fig:TMD}.
}
\end{figure}

The differential cross sections, corrected for the branching ratio to the $\pi^0\pi^0$ final state, are shown in Fig.~\ref{fig:xsec} as a function of $-t$ for four ranges of $E_\gamma$ (only statistical uncertainties are plotted). In general, the cross sections decrease with increasing beam energy, having the same dependence on $-t$, with a maximum at $-t = 0.35$~GeV$^2$. Even though the bin sizes in $E_\gamma$ are smaller than for the $f_2(1270)$ measurement of the 2009 CLAS data from the $\pi^+\pi^- p$ final state \cite{Marco}, the present cross sections are much more precise due to the lack of background from $\rho$-decay. In comparison with the cross sections for $f_2(1270)$ extracted \cite{mathieu} from the $D$-wave component of a PWA fit to the 2009 data, the present cross sections are larger.  However, that $D$-wave strength had a large uncertainty, due to the method of using a PWA fit in the presence of a large background from the $\rho$-meson decay, whereas the present results have a large signal on a small background.

The cross sections of Fig.~\ref{fig:xsec} are compared with theory predictions from model A of Xie and Oset \cite{Xie}, described above, with one free parameter (the $\rho$-$\rho$ coupling, which is fixed from other data).  In particular, these are the predictions of model A in Ref.~\cite{Xie}, but calculated for the incident photon beam energies of the present data.  Although that model compared well with the experimental results of Ref.~\cite{Marco}, using the $D$-wave strength described above (and for a different range of beam energy), it does not agree with the present results.  This suggests that a more sophisticated theoretical model is necessary. 

\begin{figure}[tpb]
\includegraphics[scale=0.45]{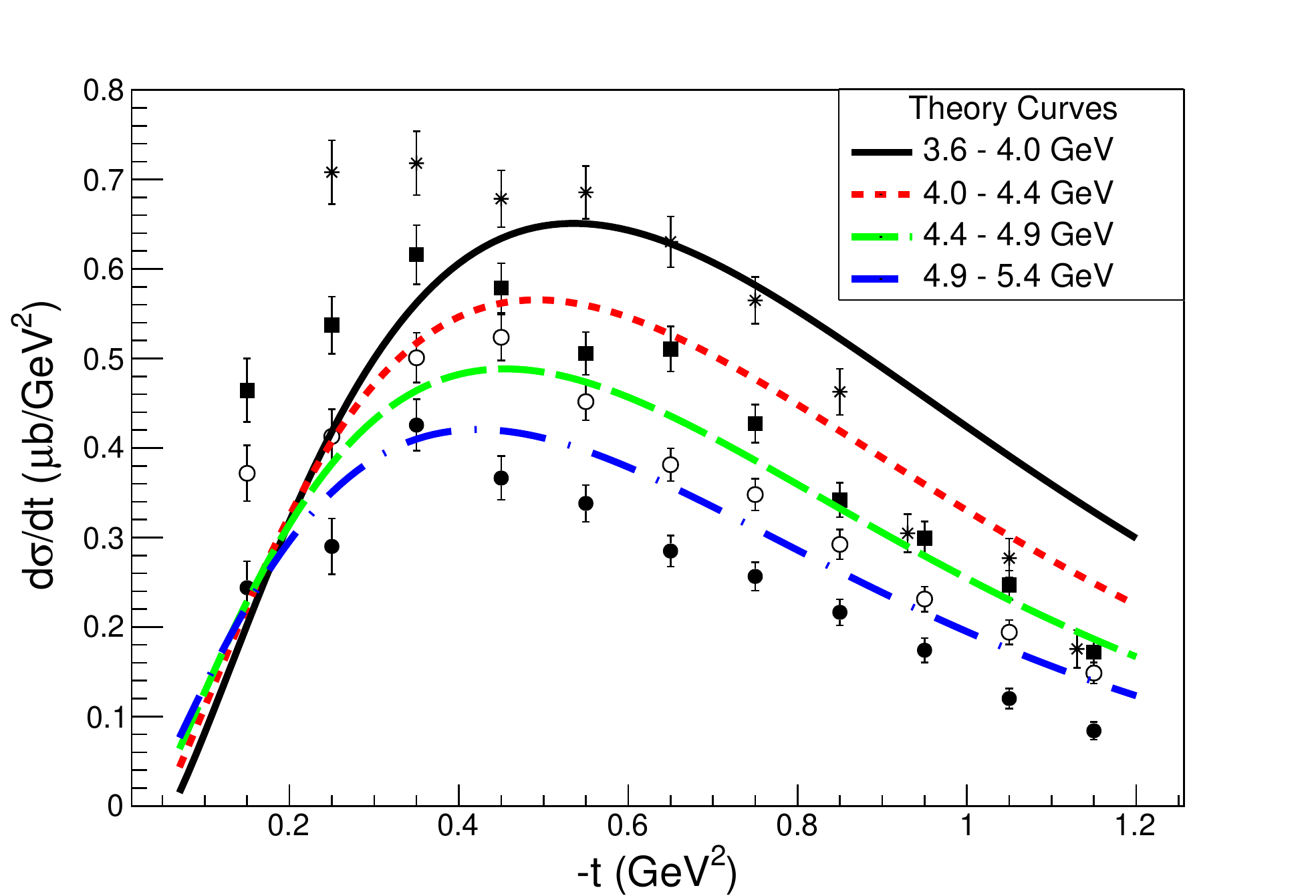}
\caption{\label{fig:TMD} Same as Fig.~\ref{fig:xsec}, except the curves are for the tensor meson dominance (TMD) model of Ref. \protect \cite{mathieu}. The curves have been scaled down (multiplied by a factor of 0.6) to keep the $y$-axis range fixed.  See also the legend of Fig.~\ref{fig:xsec}.}
\end{figure}

In Ref.~\cite{mathieu}, two tensor meson photoproduction models have been developed. They differ by the helicity structure of the photon-tensor meson vertex. In the minimal model, the tensor meson interacts via a point-like interaction with the photon, similar to the models of Refs.~\cite{Molina,Xie}, resulting in curves very similar to Fig.~\ref{fig:xsec}. In the tensor meson dominance (TMD) model, the tensor meson couples to a vector field via the stress-energy tensor. The presence of a derivative in this latter interaction implies a vanishing of the cross section in the forward direction ($t \sim -0.1$~GeV$^2$). 
For each model, the two free parameters, the strength of the vector and axial-vector exchange contributions, have been determined from a recent extraction of the $a_2(1320)$ differential cross section~\cite{AndreaEtaPi}. The predictions of the TMD model for the $f_2(1270)$ differential cross sections shown in Fig.~\ref{fig:TMD} (scaled by a factor of 0.6 for ease of comparison) are calculated by using isospin relations between the two tensor mesons.
Note that the minimal model is dominated by axial-vector exchanges and displays a milder energy dependence than the TMD model, and so the minimal model shows a non-vanishing cross section in the forward direction. 
The TMD model overestimates the data by roughly 40\%.  However, the normalization of the effective coupling constants in the TMD model was determined by comparison with data on $a_2(1320)$ photoproduction~\cite{AndreaEtaPi}, so these model parameters can be fixed by the experimental results.
These new data thus call for a global theoretical analysis of both $a_2(1320)$ and $f_2(1270)$ photoproduction. 
At present, the energy and $t$-dependence of the CLAS data, shown in Fig.~\ref{fig:xsec}, are more compatible with the TMD model and strongly suggest the dominance of vector exchanges, whose contribution vanishes in the forward direction. 

In summary, we have measured for the first time the reaction $\gamma p \to \pi^0 \pi^0 p$ at small four-momentum transfer $t$ and extracted differential cross sections for the $f_2(1270) p$ final state over four bins in photon beam energy. The results show an increase in the cross sections from $t_{min}$ up to $-t \sim 0.35$~GeV$^2$, which then falls linearly up to $-t = 1.2$~GeV$^2$.
The $t$-dependence disagrees with predictions from the model of Xie and Oset \cite{Xie}, where the $f_2(1270)$ is described as a dynamically generated resonance from the attraction of two $\rho$-mesons. The data agree better with the tensor meson 
dominance model of Ref. \cite{mathieu}, which includes both vector and axial-vector exchange to the $f_2(1270)$, assuming a quark-model structure (a $q\bar q$ pair with quantum numbers $S=1$ and $L=1$, coupled to $J=2$).  Further theoretical studies, which include both the present results and additional data on the $a_2(1320)$, are needed to more fully understand the photoproduction mechanism and hence the internal structure of the $f_2(1270)$ meson.

More experimental information on $f_2(1270)$ photoproduction is also possible.  The GlueX and CLAS12 detectors at Jefferson Lab can measure the same reaction studied here, but using linear polarization and at higher photon energies.  In addition, the CLAS measurements could be extended by utilizing circular polarization of the photon beam, which would provide more information about the reaction mechanism.  For now, the present results are a significant step forward, providing the first high-precision cross sections with small bins in $t$, which clearly distinguish between theoretical models based on vector and axial-vector meson exchange.

\begin{acknowledgments}
The authors acknowledge the staff of the Accelerator and Physics Divisions at the Thomas Jefferson National Accelerator Facility who made this experiment possible. This work was supported in part by the Chilean Comisi\'on Nacional de Investigaci\'on Cient\'ifica y Tecnológica (CONICYT), by CONICYT PIA Grant No. ACT1413, the Italian Istituto Nazionale di Fisica Nucleare, the French Centre National de la Recherche Scientifique, the French Commissariat \'a l’Energie Atomique, the United Kingdom Science and Technology Facilities Council (STFC), the Scottish Universities Physics Alliance (SUPA), the National Research Foundation of Korea, and the U.S.~National Science Foundation.
V.M. acknowledges support from the Community of Madrid through the Programa de Atracci\'on de Talento Investigador 2018-T1/TIC-10313 and from the Spanish national grant PID2019-106080GB-C21.
The Southeastern Universities Research Association operates the Thomas Jefferson National Accelerator Facility for the United States Department of Energy under Contract No. DE- AC05-06OR23177.
\end{acknowledgments}


\begin{thebibliography}{99}

  \bibitem{Tanabashi} 
  M.~Tanabashi {\it et al.} [Particle Data Group],
  Phys.\ Rev.\ D {\bf 98}, no. 3, 030001 (2018).

  \bibitem{Molina}
  R.~Molina {\it et al.}, Phys. Rev. D {\bf 78},114018 (2009).
  
  \bibitem{Xie}
  J-J. Xie and E. Oset, Eur. Phys. J. A {\bf 51}, 111 (2015).
  
  \bibitem{Marco}
  M. Battaglieri {\it et al.} [CLAS Collaboration], Phys. Rev. D {\bf 80}, 072005 (2009).
  
  \bibitem{Yu}
  Q-X. Shen and H.Yu, Phys. Rev. D {\bf 40}, 1517 (1989).
  
  \bibitem{DeMinLi}
  De Min Li {\it et al.}, J. Phys. G {\bf 27}, 807 (2001).
  
  \bibitem{Giacosa}
  F. Giacosa {\it et al.}, Phys. Rev. D {\bf 72}, 114021 (2005).
 
  \bibitem{mecking}
  B.A.~Mecking {\it et al.},
  Nucl. Instrum. Meth. Phys. Res. A {\bf 503}, 513 (2003).
  
  \bibitem{Clifft} 
  R.~W.~Clifft {\it et al.},
  Phys.\ Lett.\  {\bf 64B}, 213 (1976).

  \bibitem{mathieu} V. Mathieu {\it et al.}, Phys. Rev. D {\bf 120}, 014003 (2020).

  \bibitem{tagger}
  D.I. Sober {\it et al.},
  Nucl. Instrum. Meth. A {\bf 440}, 263 (2000).
  
  \bibitem{g12}
  Z. Akbar {\it et al.} [CLAS Collaboration], CLAS Note 2017-002,
  https://misportal.jlab.org/ul/Physics/Hall-B/clas/

    \bibitem{AndreaEtaPi}
  A. Celentano {\it et al.} [CLAS Collaboration], Phys. Rev. C {\bf 102}, 032201 (2020).

  \bibitem{clasNotes} M. Carver {\it et al.}, CLAS Note 2020-002, {\it ibid.}
  
  \bibitem{geant3}
  R. Brun {\it et al.},
  CERN Report No. CERN-DD-EE-84-1 (1987).
  
  \bibitem{kunkel} M.C. Kunkel {\it et al.} [CLAS Collaboration], 
  Phys. Rev. C 98, 015207 (2018).
  
 
    
\end{thebibliography}

\end{document}